# Fracture Characteristics of Rare-earth Phosphate under Molten Calcium-Magnesium Aluminosilicate Corrosion

Subrato Sarkar[1], Rahul Rahul[2,3,*], Bishnu Pada Majee[4], Keith Bryce[1], Lucy Zhang[1], Liping Huang[4], Jie Lian[1,4], Suvranu De[2,5]

[1]Dept. of Mechanical, Aerospace and Nuclear Engineering, Rensselaer Polytechnic Institute, Troy, NY, USA
[2]Center for Modeling, Simulation and Imaging in Medicine, Rensselaer Polytechnic Institute, Troy, NY, USA
[3]Department of Biomedical Engineering, Rensselaer Polytechnic Institute, Troy, NY, USA
[4]Department of Material Science and Engineering, Rensselaer Polytechnic Institute, Troy, NY, USA
[5]FAMU-FSU College of Engineering, Tallahassee, FL, USA

**ABSTRACT**

The fracture characteristics of $LuPO_4$ rare-earth phosphate environmental barrier coating (EBC) material under molten calcium-magnesium aluminosilicate (CMAS) corrosion is quantified. EBCs are crucial for protecting SiC-based ceramic matrix composite components in the hot section of gas turbine engines. Recent research has highlighted the potential of rare-earth phosphates as better EBC materials than third-generation rare-earth silicates for CMAS corrosion resistance. However, the fracture of EBCs under CMAS corrosion during service remains a significant concern. This work investigates the fracture behavior of $LuPO_4$ using mesoscale simulation and experiments. The model uses micrographs taken from fabricated EBC samples for mesoscale fracture simulations. The simulation results are compared with experimental fracture toughness data and validated using statistical tests ($p<0.01$). The simulation results and experimental observations demonstrate that $LuPO_4$ may exhibit higher fracture resistance than $Lu_2SiO_5$ rare-earth silicate under similar CMAS corrosion conditions, offering potential insights for future EBC design and development.

**INTRODUCTION**

Environmental barrier coatings (EBCs) are used to protect silicon carbide (SiC) based ceramic matrix composites (CMC) from high-temperature corrosion in the hot-section of gas turbine engines [1]. Recently, rare-earth phosphates are shown to have significant advantages over third-generation rare-earth silicates for molten calcium-magnesium aluminosilicate (CMAS) corrosion resistance [2], [3], [4], [5]. However, it is unclear if the advantages of corrosion resistance translate into improved fracture resistance for rare-earth phosphates compared to silicates. Improving fracture resistance is crucial for EBCs to avoid microcracking, which is one of the primary reasons for spallation and eventual failure [6], [7], [8]. This work investigates the fracture behavior of rare-earth phosphate in comparison to rare-earth silicate EBCs under CMAS corrosion using mesoscale simulations and experiments.

---

* Corresponding author, Tel: +1 (518) 276-6707; Email: rahul@rpi.edu

Silicon carbide (SiC) based ceramic matrix composites (CMC) have been proven to be a suitable replacement for Ni-based superalloys used in the hot section components of gas turbine engines due to their stable high-temperature performance [1], [9]. However, when SiC-based CMCs are subject to high-temperature gas streams at high speed within the gas turbines, they undergo different reactions and surface recession (corrosion) [10]. One of the leading causes of corrosion in CMCs is the reaction and penetration of molten CMAS formed from ingested debris, *i.e.*, sand, dust or volcanic ash [11]. To protect CMCs from corrosion, environmental barrier coatings (EBCs) are used. EBCs are designed to be mechanically and chemically stable at high temperatures, providing a protective barrier between the high-temperature gas stream and CMCs [12].

In the past several decades, numerous materials have been developed for EBC applications. These materials are broadly categorized into three generations based on their increasing high-temperature capability realized by introducing different high-temperature resistant compounds. The first generation EBCs contain mullite ($Al_2O_3 \cdot SiO_2$), the second generation contains BSAS ($BaO \cdot SrO \cdot Al_2O_3 \cdot SiO_2$) and the third generation contains rare-earth (RE) silicates ($RE_2Si_2O_7/RE_2SiO_5$) in their composition [13]. Among these, the third generation EBCs based on rare-earth disilicates are more durable and offer better protection against CMAS corrosion than earlier generations [12]. However, rare-earth disilicates still corrode rapidly when exposed to molten CMAS at temperatures of 1300°C and above, as the CMAS melt penetrates along the grain boundaries, forming large cracks and severe swelling [14], [15]. Recently, this shortcoming has been addressed by replacing disilicates with phosphates, which form a continuous and dense reaction layer that halts CMAS penetration into the bulk pellet and limits corrosion [4], [5]. With less CMAS corrosion, rare-earth phosphates are a promising EBC material with durable high-temperature performance for hot sections of gas turbine engines.

This work focuses on quantifying the effects of CMAS corrosion on the fracture strength of EBCs using mesoscale simulations. During CMAS corrosion, molten CMAS penetrates the EBC microstructure with randomly oriented ceramic grains and causes structural changes [5], [14]. The penetration into the microstructure is driven by total energy reduction through the formation of two CMAS-ceramic interfaces (low energy) replacing one ceramic-ceramic interface (high-energy grain boundary) [16]. The penetrated CMAS forms a brittle and glassy phase along ceramic grain boundaries that degrade the fracture strength of EBC [12], [14]. However, it is crucial to analyze and quantify the extent of degradation caused by penetrated CMAS, which might help understand the fracture performance of EBCs under CMAS corrosion. Hence, the structural degradation of EBC microstructures is analyzed by estimating fracture resistance for different CMAS penetration levels based on experimental observations.

Among the available computational fracture models for polycrystalline materials, phase field and cohesive zone-based models are the most frequently used [17], [18], [19], [20], [21], [22]. This work uses a mesoscale model based on the cohesive finite element model (CFEM) for fracture simulation and estimating fracture resistance of EBC microstructures [23], [24]. The cohesive finite element model (CFEM) has been used in the past to identify microstructural fracture patterns, *i.e.*, intergranular/transgranular fracture, in metals and ceramics [24], [25], [26], [27] and for estimating fracture resistance of microstructures [23], [26], [28], [29].

This work uses the CFEM to simulate fracture through EBC microstructures before and after CMAS penetration [23]. The microstructures used in the models consist of randomly oriented ceramic grains and their grain boundaries. The CMAS penetration is modeled by adding CMAS along the grain boundaries [14]. The fracture is tracked using cohesive elements inserted along all element edges throughout the model, as described in the Methods section. The cohesive elements are assigned different traction-separation laws based on the material or interface they represent (see Methods section). Complex fracture patterns arising due to anisotropic ceramic grains, orientation-dependent grain boundaries, and the presence of CMAS are accounted using different material properties and traction separation laws. The fracture resistance of the EBC is estimated by computing *J*-integral [30] and fracture toughness ($K_{\text{Ic}}$). The computed $K_{\text{Ic}}$ is validated with the experimental $K_{\text{Ic}}$ measured from indentation fracture tests on EBC samples before and after CMAS penetration. Additionally, statistical tests are performed to asses similarities or differences between the simulation results and experimental measurements (Supplementary Figure 3 and Tables 1-3). A detailed description of the CFEM, *J*-integral, and $K_{\text{Ic}}$ computation and measurement of experimental fracture toughness is provided in the Methods section. This work uses a combination of simulations and experiments to get meaningful insights into the fracture behavior of the EBC materials under CMAS penetration, as depicted in Figure 1. Based on the observations in this work, the approach comprising simulations and experiments may be a promising way to obtain actionable insights into the behavior of new materials for whom extensive testing is infeasible.

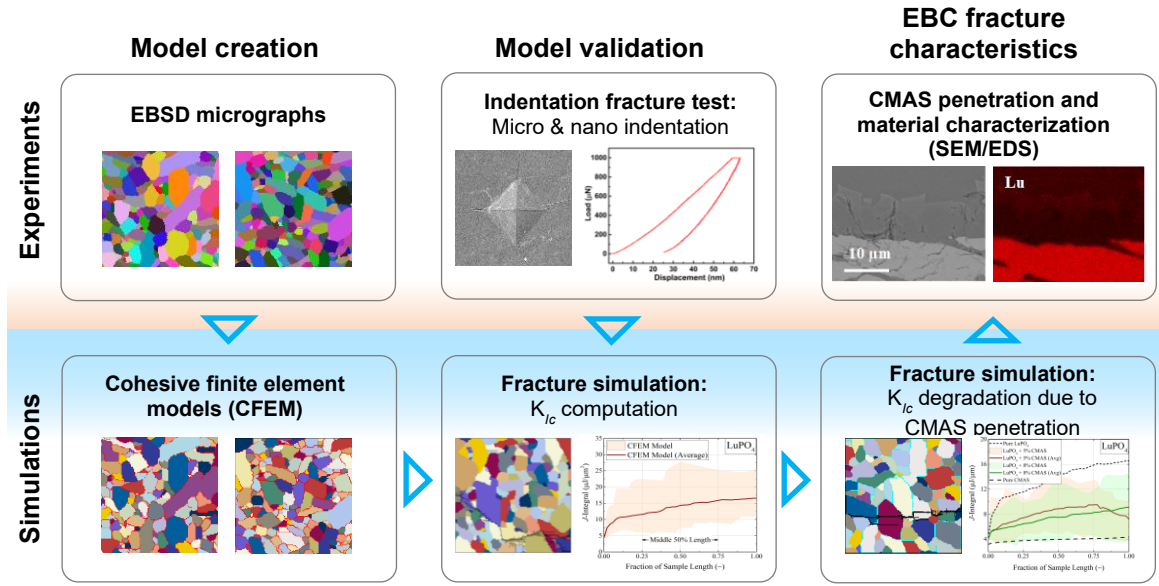

**Figure 1: A schematic flow diagram showing the combined experiments and simulation approach adopted in this work.** The cohesive finite element models (CFEM) are formulated using experimentally obtained electron backscattered diffraction (EBSD) micrographs of EBC samples. The CFEM is used for fracture simulations and fracture toughness ($K_{Ic}$) computation. The computed $K_{Ic}$ is validated with the experimental $K_{Ic}$ obtained using micro and nanoindentation tests [44] performed on EBC samples before and after CMAS penetration. The validated model is then used to investigate the fracture characteristics of EBCs with CMAS penetration. The microstructural characterization of EBC samples is carried out using scanning electron microscopy (SEM) and energy-dispersive X-ray spectroscopy (EDS). The computed $K_{Ic}$ values and microstructural characterization are used to compare the fracture resistance of rare-earth phosphate and silicate EBCs.

## RESULTS

### Model Validation

Among the available rare-earth (RE) metals for EBC applications, lutetium (Lu) is chosen in this work as lutetium phosphate ($LuPO_4$) and lutetium silicate ($Lu_2SiO_5$). They exhibit better CMAS resistance and thermal expansion compatibility as EBC topcoats than other RE phosphates and silicates [5], [12], [31]. Unlike single-layered coating designs used in phosphate EBCs ([4], [32]), multi-layered coating designs are generally adopted in silicate EBC applications [12]. The multi-layered design consists of $Lu_2SiO_5$ (monosilicate) top coat and an interlayer of $Lu_2Si_2O_7$ (disilicate) below the top coat but above the bond coat. As a result, $Lu_2SiO_5$ (monosilicate), being the top coat, serves as the first line of defense against CMAS in silicate EBC and is thus chosen for analysis in this work.

### Validation of CFEM without CMAS penetration

The mesoscale simulations are carried out for $LuPO_4$ and $Lu_2SiO_5$ using eight different microstructure (EBSD) images taken from fabricated EBC samples (Supplementary Figures 1-2). A schematic representation of the CFEM model with boundary conditions is shown in Figure 7 (Methods section). A total of 32 simulations are carried out by initiating fracture on each of the four sides of the experimentally obtained microstructures. The simulation results for the LuPO$_4$ and Lu$_2$SiO$_5$ without CMAS penetration are shown in Figure 2. The pure EBCs exhibit a tortuous fracture propagation (Figure 2a-b), which could be attributed to the material anisotropy and difference in the fracture strengths of grains and grain boundaries. The anisotropy and inhomogeneity in the microstructure cause the crack path to pass through grains (transgranular fracture) and along grain boundaries (intergranular fracture). The crack path is quantified using a crack path fraction parameter defined as fractions of crack path inside grains, *i.e.,* $H_{\text{GR}} = \frac{L_{\text{GR}}}{L_{\text{GB}}+L_{\text{GR}}}$, and fraction of crack path along the grain boundaries, *i.e.*, $H_{\text{GB}} = \frac{L_{\text{GB}}}{L_{\text{GR}}+L_{\text{GB}}}$, where $L_{\text{GB}}$ and $L_{\text{GR}}$ are crack lengths within grains and along grain boundaries, respectively [23]. The mean values of crack path fraction are found to be around 0.6 and 0.4 for the grains and grain boundaries, respectively, for both the LuPO$_4$ and Lu$_2$SiO$_5$ EBCs (Figure 2c). The similarity of the crack path fractions indicates the dominant effect of the microstructure on the fracture behaviour of EBCs, where same set of microstructures are used for the fracture simulations of both EBCs.

The model is validated by comparing fracture toughness $K_{\text{Ic}}$ of the EBCs obtained from CFEM simulations to that of the experimentally measured values. The $K_{\text{Ic}}$ is estimated from the *J*-integral values obtained from simulations using the relation outlined in the Methods section. The *J*-integral values for LuPO$_4$ and Lu$_2$SiO$_5$ EBCs with evolving crack length are shown in Figure 2d and 2e, respectively. The noticeable variation in the *J*-integral values for different microstructures could be attributed to the tortuosity of the crack path (Figures 2a and 2b) [33]. Hence, the mean *J*-integral values for the middle 50% of the sample length are used to estimate the $K_{\text{Ic}}$ of the samples mitigating the edge-effects. The estimated $K_{\text{Ic}}$ values for the LuPO$_4$, Lu$_2$SiO$_5$, and CMAS samples are compared with that of the experimentally measured values in Figure 2f [25], [34]. Notice that the $K_{\text{Ic}}$ for homogenous and isotropic CMAS is almost constant with zero variance (Figure 2f).

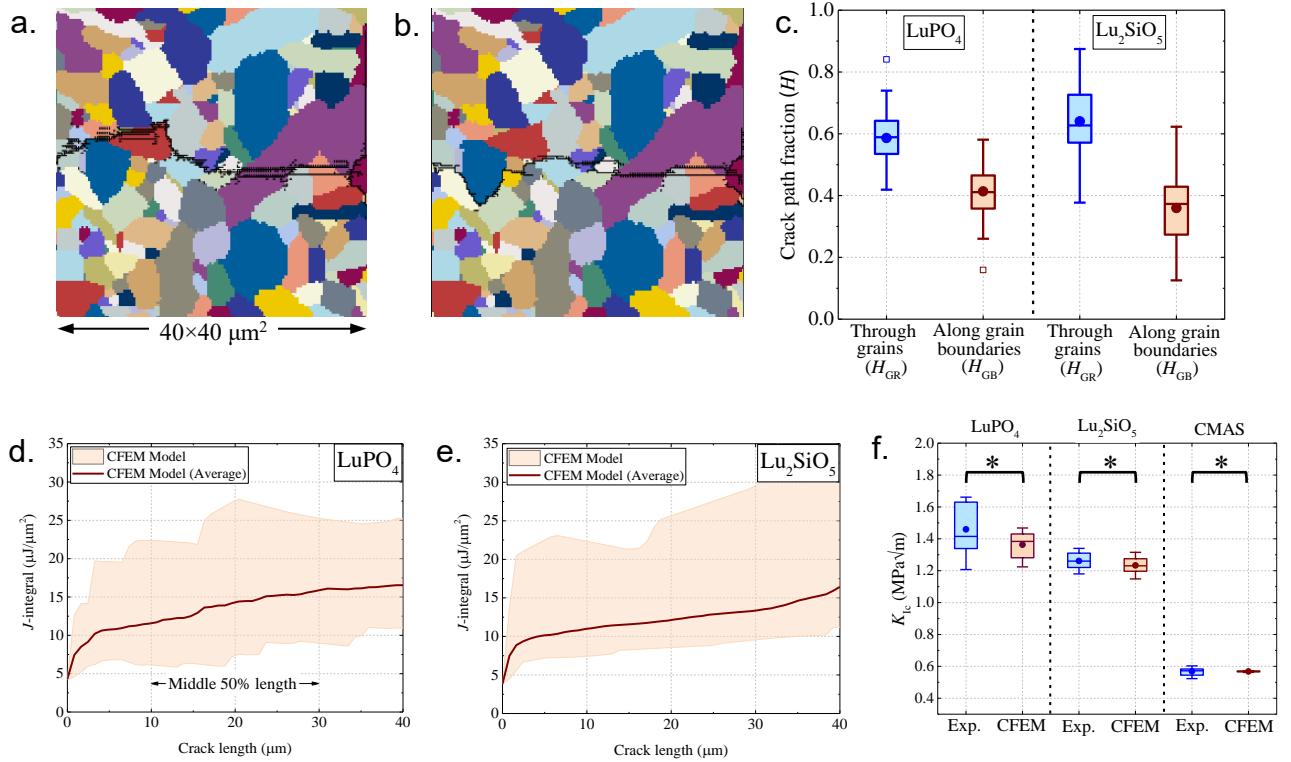

**Figure 2: Simulation results from cohesive finite element model (CFEM) for $LuPO_4$ and $Lu_2SiO_5$ EBCs, and CMAS.** Fracture patterns in the **(a)** $LuPO_4$ and **(b)** $Lu_2SiO_5$ microstructures through grains and along grain boundaries **(c)** The box plots show crack path fractions for EBCs. The *J*-integral values for **(d)** $LuPO_4$ and **(e)** $Lu_2SiO_5$ EBCs with evolving crack length. The variations in *J*-integral values for all the microstructures are shown as light-colored bands and their average is the dark-colored curve. Panel **(f)** compares fracture toughness ($K_{Ic}$) obtained from CFEM simulations with that of experimentally obtained values. The '*' indicates similarity of the data from experiments and CFEM simulations based on two-sample *t*-test and Wilcoxon rank-sum test, see supplementary data for details.

For validation in Figure 2f, 10, 11 and 15 experimental measurements of indentation fracture toughness ($K_{Ic}$) are used for $LuPO_4$, $Lu_2SiO_5$ and CMAS respectively, with 32 values computed for each material from the simulations. The similarity between computed and experimentally measured $K_{Ic}$ values is assessed using statistical tests based on a two-sample *t*-test and Wilcoxon rank-sum test with $p < 0.01$ (Supplementary Table 2). The tests show that the $K_{Ic}$ values estimated from simulations are indeed statistically similar to that of experimentally measured values, thus validating simulation results for pure EBCs and CMAS. The observations demonstrate the ability of the CFEM to simulate fracture accurately in pure $LuPO_4$, $Lu_2SiO_5$ and CMAS.

**Validation of CFEM with CMAS penetration**

The CFEM is further validated to describe the fracture behavior of CMAS-penetrated EBCs. Two levels of CMAS penetration, *i.e.*, 5% and 8% by volume, are introduced along the grain boundaries within the microstructures obtained from the fabricated EBC samples. These two levels are chosen based on experimental observations in this work and literature [35] that indicate CMAS penetration along the grain boundaries within the in the range of 0 – 15% by volume. The CFEM simulations are carried out using boundary conditions and material properties presented in the Methods section. The $K_{Ic}$ values estimated from the *J*-integral values are compared with the experimental $K_{Ic}$ of LuPO$_4$ and Lu$_2$SiO$_5$ with CMAS (5% and 8% by volume) measured from indentation fracture testing.

    The simulation results for 5% and 8% CMAS-penetrated LuPO$_4$ and Lu$_2$SiO$_5$ EBC samples are shown in Figure 3. A total of 32 simulations are carried out for each of the EBCs with CMAS penetration by initiating fracture on each of the four sides of the microstructure obtained from eight EBSD images of EBCs. The tortuous fracture patterns observed in a LuPO$_4$ microstructure at 5% and 8% CMAS penetration are shown in Figures 3a through 3d. Fracture patterns show that the crack predominantly propagates through the penetrated CMAS regions. This behavior is expected as the fracture strength of CMAS and CMAS-grain interfaces are considered lower than the grains and grain boundaries (Table 2). The crack path fractions for the CMAS-penetrated microstructures are shown in Figure 3e. The crack path fractions are defined as fractions of the crack path inside grains, *i.e.*, $H_{GR} = \frac{L_{GR}}{L_{GR}+L_{GB}+L_{CM}+L_{CI}}$, along the grain boundaries, *i.e.*, $H_{GB} = \frac{L_{GB}}{L_{GR}+L_{GB}+L_{CM}+L_{CI}}$, inside CMAS, *i.e.*, $H_{CM} = \frac{L_{CM}}{L_{GR}+L_{GB}+L_{CM}+L_{CI}}$, and along grain-CMAS interface, *i.e.*, $H_{CI} = \frac{L_{CI}}{L_{GR}+L_{GB}+L_{CM}+L_{CI}}$, where $L_{GR}$, $L_{GB}$, $L_{CM}$ and $L_{CI}$ are crack lengths within grains, along grain boundaries, within penetrated CMAS, and grain-CMAS interface, respectively. The crack path fractions indicate that increasing CMAS penetration from 5% to 8% causes increased crack propagation through CMAS-grain interfaces (CI) in both LuPO$_4$ and Lu$_2$SiO$_5$ samples. Conversely, cracks passing through grains (GR) and grain boundaries (GB) decrease.

    The *J*-integral values computed using CFEM simulations for CMAS-penetrated LuPO$_4$ and Lu$_2$SiO$_5$ are shown in Figures 3f and 3g as a function of evolving crack length. The mean value of *J*-integral for both 5% and 8% CMAS penetration lies below the *J*-integral value of pure LuPO$_4$/Lu$_2$SiO$_5$, indicating degradation due to the presence of CMAS in the microstructure.

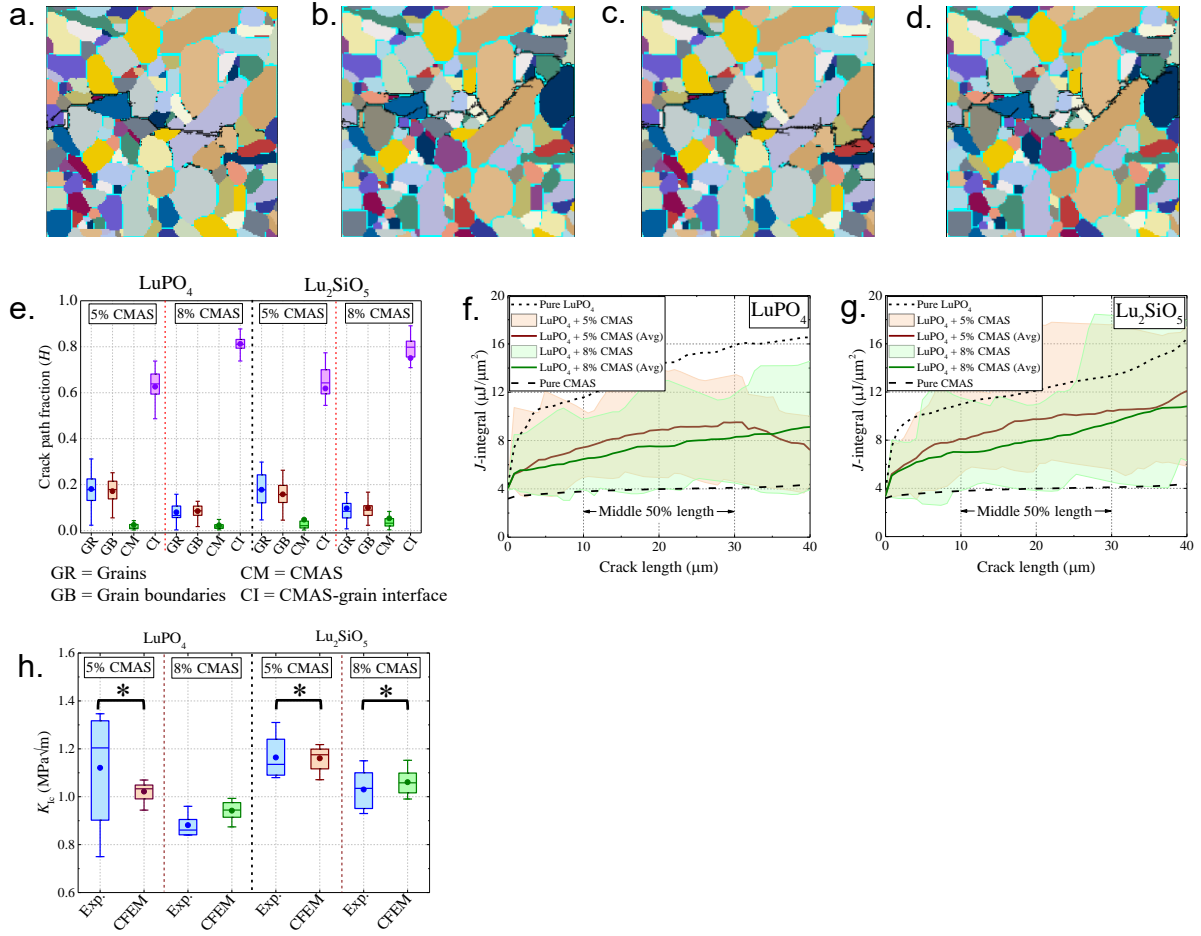

**Figure 3: Simulation results from cohesive finite element model (CFEM) for 5% and 8% CMAS-penetrated LuPO$_4$ and Lu$_2$SiO$_5$ EBCs.** Panels **(a)-(b)** and **(c)-(d)** show fracture patterns in the 5% and 8% CMAS-penetrated LuPO$_4$ and Lu$_2$SiO$_5$ EBC microstructures respectively. The box plots in panel **(e)** shows crack path fractions for EBCs with CMAS penetration. The $J$-integral values for **(f)** LuPO$_4$ and **(g)** Lu$_2$SiO$_5$ with and without CMAS penetration are compared with that of the pure CMAS with evolving crack length. The variations in $J$-integral values for all the microstructures are shown as light-colored bands and their average is presented with the dark-colored curve. Panel **(h)** compares fracture toughness ($K_{Ic}$) obtained from CFEM simulations with that of experimentally obtained values. The '*' indicates similarity of the data from experiments and CFEM simulations based on two-sample $t$-test and Wilcoxon rank-sum test, see supplementary data for details.

The model is validated by comparing fracture toughness $K_{Ic}$ of CMAS-penetrated EBCs with that of the experimentally measured values. The $J$-integral from the middle 50% of the simulation length is used to estimate $K_{Ic}$ to mitigate the edge effects. The distribution of $K_{Ic}$ for EBCs with the 5% and 8% CMAS penetration is presented using box plots in Figure 3h. For validation in Figure 3h, 06 and 05 values of experimentally measured indentation fracture toughness ($K_{Ic}$) are used for LuPO$_4$ with 5% and 8% CMAS, respectively, with 32 values for each material from the simulations. While for Lu$_2$SiO$_5$, 10 values of experimental $K_{Ic}$ are used for both 5% and 8% CMAS penetrated microstructures. The $K_{Ic}$ values obtained from simulations and experiments are compared and their similarity is assessed using statistical tests based on a two-sample *t*-test and Wilcoxon rank-sum test with $p < 0.01$. The tests in Figure 3f show that the estimated $K_{Ic}$ is statistically similar to that of experimentally measured values for three out of four cases (Supplementary Table 3). The fracture toughness for LuPO$_4$ with 8% CMAS penetration does not show statistical similarity with the experimental values (Figure 3f). This could be due to insufficient experimental observations needed to test the statistical similarity between measured and computed $K_{Ic}$. Nonetheless, the comparison shows reasonable agreement for both LuPO$_4$ and Lu$_2$SiO$_5$ in most cases. In addition, it is seen that an increase in CMAS penetration from 5% to 8% led to a decrease in $K_{Ic}$, indicating higher degradation in fracture resistance with increasing CMAS penetration.

The statistical tests further reinforce the efficacy of CFEM in predicting the fracture behavior of CMAS-penetrated EBCs.

**Fracture Resistance of LuPO$_4$ and Lu$_2$SiO$_5$ with CMAS Penetration**

The validated CFEM is employed to investigate the fracture behavior of LuPO$_4$ and Lu$_2$SiO$_5$ EBCs with increasing CMAS penetration. Based on observations from CMAS reaction tests on EBC samples and literature [35], five levels of CMAS penetration are considered for fracture simulations: 3%, 5%, 8%, 11% and 15% by volume. Similar to the validation studies, a total of 32 simulations are carried out for each penetration level, and their results are shown in Figure 4.

The tortuous fracture patterns obtained using CFEM before and after CMAS penetration for both LuPO$_4$ and Lu$_2$SiO$_5$ EBC microstructures are shown in Figures 4a and 4b, respectively. The tortuosity of fracture in EBCs without CMAS penetration (Pure EBC) is due to the anisotropy of grains. On the other hand, crack path tortuosity after CMAS penetration can be attributed to the presence of CMAS along the grain boundaries and its lower fracture strength than grains and grain boundaries. The lower fracture strength of CMAS and its interfaces causes the fracture to pass through CMAS regions in the CMAS-penetrated microstructures (Figures 4a and 4b). It is observed that the fraction of total crack length passing through grains and grain boundaries decreases with increasing CMAS penetration, while the fraction of crack length passing through CMAS and CMAS-grain interface increases at the same time (Figures 4c and 4d). Moreover, a steep decrease in the fraction of crack path passing through grains and grain boundaries at 3% CMAS penetration compared to pure LuPO$_4$ and Lu$_2$SiO$_5$ indicates a significant change in fracture behavior caused by CMAS.

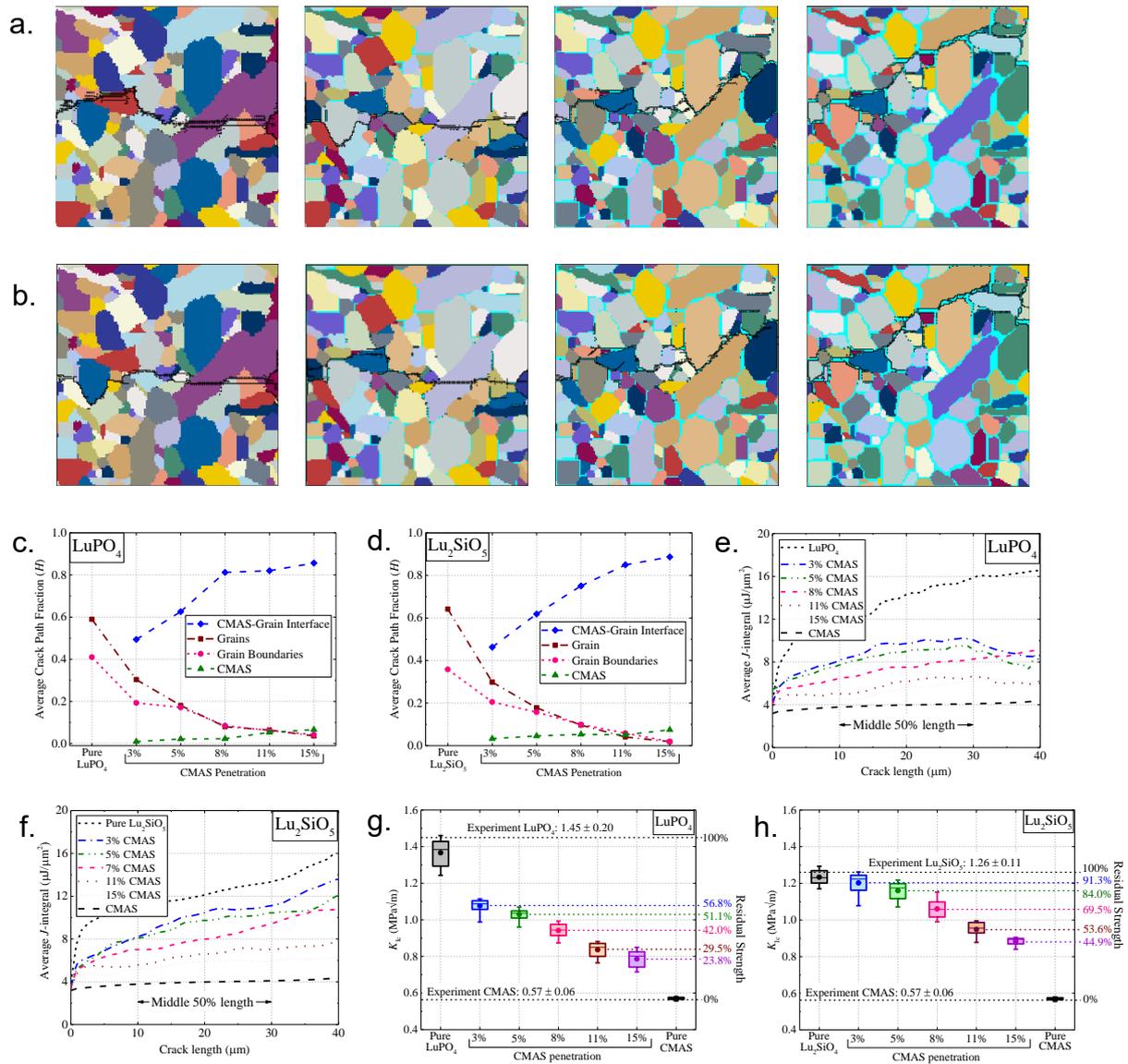

**Figure 4: Simulation results obtained using cohesive finite element model (CFEM) for LuPO$_4$ and Lu$_2$SiO$_5$ microstructures at different levels of CMAS penetration.** The fracture patterns in **(a)** LuPO$_4$ and **(b)** Lu$_2$SiO$_5$ EBC microstructures are shown with 0% CMAS (pure), 3% CMAS, 8% CMAS and 15% CMAS penetration. Panels **(c)** and **(d)** show average crack path fractions before and after CMAS penetration in LuPO$_4$ and Lu$_2$SiO$_5$, respectively. The *J*-integral values for **(e)** LuPO$_4$ and **(f)** Lu$_2$SiO$_5$ before and after CMAS penetration are compared with that of the pure CMAS with evolving crack length. Panels **(g)** and **(h)** compare fracture toughness ($K_{Ic}$) and residual strength before and after CMAS penetration in LuPO$_4$ and Lu$_2$SiO$_5$, respectively.

The average *J*-integral curves obtained from simulation for LuPO$_4$ and Lu$_2$SiO$_5$ with different CMAS penetration levels are shown in Figures 4e and 4f. For reference, the average *J*-integral curves of pure LuPO$_4$, Lu$_2$SiO$_5$, and CMAS are shown as dashed curves at the top and bottom of the plots. It is observed that the average *J*-integral curves of CMAS-penetrated LuPO$_4$ and Lu$_2$SiO$_5$ microstructures progressively degrade with increasing CMAS penetration level and approach the average *J*-integral curve of CMAS. The *J*-integral values in the middle 50% sample lengths are used to estimate $K_{Ic}$. The distribution of $K_{Ic}$ values estimated from all 32 simulations performed on eight experimentally obtained microstructures are plotted as box plots for CMAS-penetrated $LuPO_4$ and $Lu_2SiO_5$ in Figures 4g and 4h, respectively. The degradation due to CMAS penetration can be seen more clearly in the box plots. The $K_{Ic}$ degradation at different levels of CMAS penetration is indicated as residual strength percentages by assuming the fracture strength of pure $LuPO_4$ and $Lu_2SiO_5$ as 100% and that of the CMAS as 0%. The degradation of $K_{Ic}$ in $LuPO_4$ is observed to be higher than in $Lu_2SiO_5$. With a further increase in CMAS penetration, the $K_{Ic}$ values degrade further and approach CMAS values. A steep drop in $K_{Ic}$ caused by low levels of CMAS penetration in $LuPO_4$ emphasizes the need for mechanisms that can limit CMAS penetration into the microstructures.

**DISCUSSION**

A cohesive finite element analysis framework is developed to investigate the effect of CMAS penetration on the fracture behavior in EBCs. The CFEM is first validated with the experimentally measured fracture toughness of pure and CMAS-penetrated EBCs. The validated model is then employed to investigate fracture characteristics of $LuPO_4$ in comparison with $Lu_2SiO_5$ with increasing CMAS penetration. The simulation results reveal that CMAS corrosion causes degradation of fracture resistance ($K_{Ic}$) in both $LuPO_4$ and $Lu_2SiO_5$ EBCs, with more degradation observed in $LuPO_4$ (Figures 4g and 4h). It is seen that before CMAS penetration, the $K_{Ic}$ of $LuPO_4$ was slightly higher than that of $Lu_2SiO_5$. However, after CMAS penetration, the $K_{Ic}$ values of $LuPO_4$ are generally lower than $Lu_2SiO_5$. The lower values of $K_{Ic}$ for $LuPO_4$ are also observed experimentally in the indentation test results shown in Figure 3h.

These observations indicate that the fracture resistance of CMAS-penetrated $LuPO_4$ is lower than that of $Lu_2SiO_5$. Interestingly, it also emphasizes that a higher fracture resistance of $LuPO_4$, in comparison with $Lu_2SiO_5$, could be achieved by eliminating or minimizing CMAS corrosion. To test this hypothesis, we have experimentally studied the CMAS corrosion in $LuPO_4$ and $Lu_2SiO_5$ by exposing them to molten CMAS at 1300 °C for 45 and 5 hours, respectively. Notice that the $LuPO_4$ sample was subjected to a 45-hour CMAS reaction to observe its CMAS corrosion resistance under more severe conditions with a longer duration.

The backscattered scanning electron microscopy (SEM) and energy dispersive X-ray spectroscopy (EDS) images of $LuPO_4$ and $Lu_2SiO_5$ after the CMAS reaction are shown in Figures 5 and 6, respectively. Figures 5a and 6a show the epoxy, CMAS, and bulk EBC pellet after CMAS reaction in $LuPO_4$ and $Lu_2SiO_5$ samples. It is observed that $Lu_2SiO_5$ delaminates after the CMAS reaction of 5 hours, whereas the LuPO$_4$ remains intact even after 45 hours. The delamination in Lu$_2$SiO$_5$ could be attributed to the mismatch in the coefficient of thermal expansion between Lu$_2$SiO$_5$ (~6.7×10$^{-6}$ in the temperature range 200°C-1350°C) and oxyapatite Ca$_2$Lu$_8$(SiO$_4$)$_6$O$_2$ phase formed in the reaction layer [36]. Further investigation is needed to achieve a mechanistic understanding of the stress distribution and delamination in the Lu$_2$SiO$_5$.

The high-resolution SEM images of the reaction zone in LuPO$_4$ and Lu$_2$SiO$_5$ samples are shown in Figures 5b and 6b, respectively. It is seen that a reaction layer is formed at the interface of CMAS and EBCs during the CMAS reaction. The average thicknesses of the reaction layer for LuPO$_4$ and Lu$_2$SiO$_5$ are measured (at 20 locations on the SEM image) to be 32.8±3.8 µm and 6.1±1.4 µm, respectively. LuPO$_4$ forms a continuous and dense reaction layer than the Lu$_2$SiO$_5$. This aligns with the observations that RE silicates form porous and non-uniform reaction layers under CMAS corrosion [37]. At the same time, the RE phosphates have been observed to form continuous and denser reaction layers [4].

The EDS maps of CMAS-LuPO$_4$ and -Lu$_2$SiO$_5$ interfaces are shown in Figures 5c and 6c, respectively. In LuPO$_4$, calcium (Ca) did not penetrate through the reaction layer. However, a noticeable penetration of Ca is seen in Lu$_2$SiO$_5$. The penetration of Ca in LuPO$_4$ is restricted, possibly due to the formation of a continuous and dense reaction layer of Ca$_8$MgLu(PO$_4$)$_7$ along the CMAS-LuPO$_4$ interface. On the other hand, in Lu$_2$SiO$_5$, a porous and non-uniform reaction layer of oxyapatite Ca$_2$Lu$_8$(SiO$_4$)$_6$O$_2$ may have allowed Ca to penetrate in bulk Lu$_2$SiO$_5$. Magnesium (Mg) and Aluminum (Al) penetration is observed to be similar in both LuPO$_4$ and Lu$_2$SiO$_5$.

The color intensity of the EDS maps suggests that the amount of Lu present in the reaction layer is less in LuPO$_4$ (Figure 5c) than in Lu$_2$SiO$_5$ (Figure 6c). This indicates that less LuPO$_4$ is needed to form a dense reaction layer compared to Lu$_2$SiO$_5$. A lower stoichiometric ratio of Lu:Ca in Ca$_8$MgLu(PO$_4$)$_7$ (at the CMAS-LuPO$_4$ reaction layer) as compared to a higher Lu:Ca ratio in Ca$_2$Lu$_8$(SiO$_4$)$_6$O$_2$ (at CMAS-Lu$_2$SiO$_5$ reaction layer) may have played a role in the amount of Lu in the reaction layer [38]. In contrast, a high concentration of Ca is required to form the Ca$_8$MgLu(PO$_4$)$_7$ interfacial phase, limiting its penetration into the EBC matrix as compared with that of Lu$_2$SiO$_5$, and with the formation of the dense reaction layer at the interface, inhibiting the penetration of CMAS.

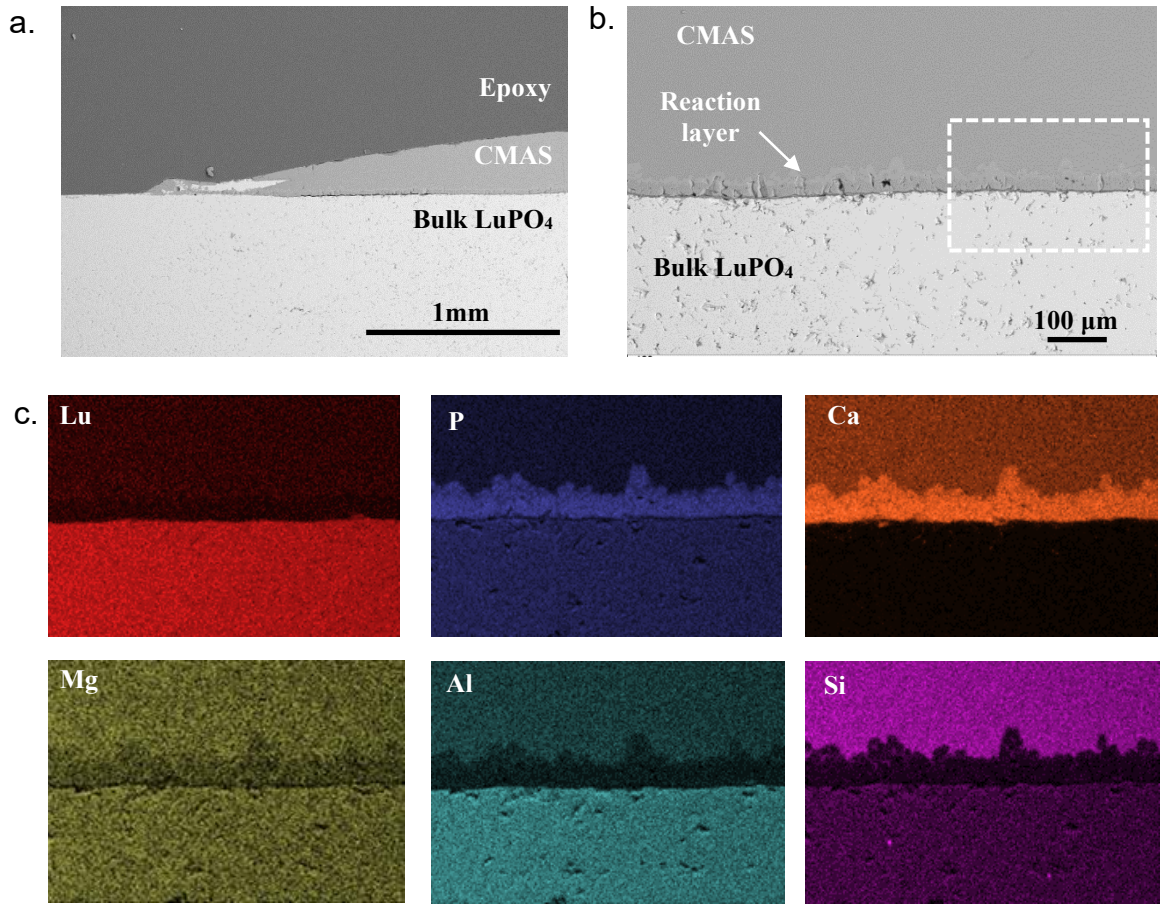

**Figure 5: SEM backscattered electron images of LuPO$_4$ interaction with molten CMAS at 1300 °C for 45 hours. (a)** Cross-section SEM image, **(b)** Higher resolution cross-section SEM image, **(c)** EDS elemental maps.

The abovementioned observations indicate that LuPO$_4$ might perform better than Lu$_2$SiO$_5$ at restricting the amount of CMAS penetrating into the bulk region by forming a dense and thicker reaction layer [32] with higher contents of Ca and less Lu. As a result, LuPO$_4$ might be able to avoid degradation of $K_{Ic}$ from CMAS penetration and retain its fracture resistance for a longer duration of time with the formation of the dense interfacial layer. The dense reaction layer was also observed in the LuPO$_4$ upon CMAS interaction at 5 hours at 1300 °C with a thickness of 16.2±3.6 μm. The greater amount of CMAS penetration in the Lu$_2$SiO$_5$ may significantly alter the thermal expansion coefficient of the matrix as well, leading to the delamination of the bulk matrix underneath the reaction layer, in contrast to the well-maintained structural integrity of the LuPO$_4$ matrix without cracking or delamination upon CMAS interaction at the same temperature but elongated duration.

The experimental observations clearly indicate that LuPO$_4$ may be able to retain a higher fracture resistance ($K_{Ic}$) due to less CMAS penetration than Lu$_2$SiO$_5$ under similar CMAS corrosion conditions.

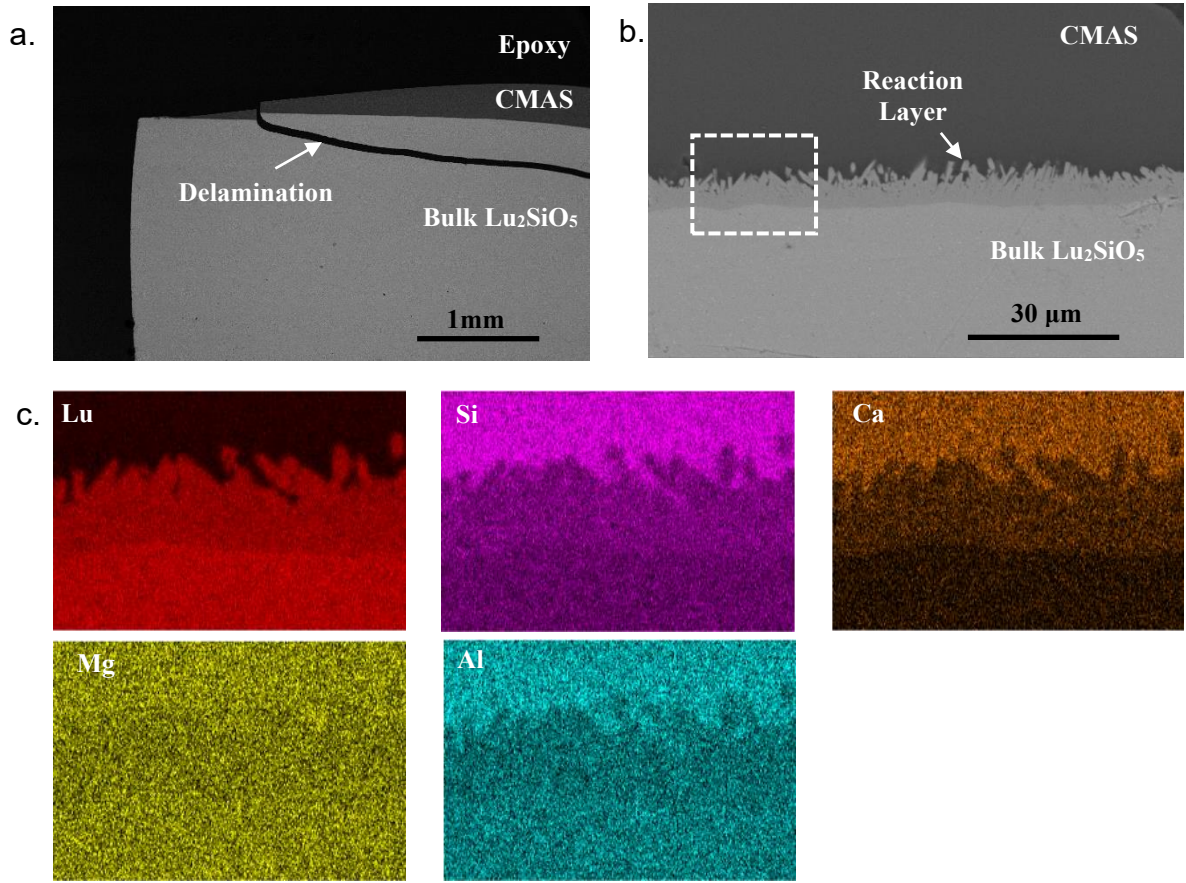

Figure 6: SEM backscattered electron images of $Lu_2SiO_5$ interaction with molten CMAS at 1300 °C for 5 hours. (a) Cross-section SEM image, (b) Higher resolution cross-section SEM image, (c) EDS elemental maps.

## METHODS

**EBC Microstructure Models**

The microstructures used in CFEM simulations are constructed from micrographs obtained using electron backscatter diffraction (EBSD). EBSD is a scanning electron microscope-based technique that scans the surface of a crystalline sample [39]. This work uses eight EBSD micrographs obtained by scanning different locations on the surface of a spark plasma sintered lutetium phosphate ($LuPO_4$) sample. The EBSD micrographs represent data as a 2D array of pixels representing multiple scanned points on the sample surface. The color of each pixel indicates a unique orientation, *i.e.*, Euler angles: $\phi_1$, $\Phi$, $\phi_2$, based on the crystal structure at a particular point on the sample surface. The scan size of each micrograph used is 40×40 sqμm with 133 × 133 pixels along the x and y direction (Supplementary Figure 1). The raw EBSD data is processed through an open-source MATLAB® (MathWorks Inc., Natick, Massachusetts, USA) toolbox MTEX [36]. While processing, the microstructure is assumed to have no porosity, and missing data in the EBSD is filled using the nearest neighbor algorithm to get a microstructure with contiguous grains [37]. After processing the EBSD data, a uniformly meshed grid of 2D plane strain quadrilateral finite elements is laid on the EBSD microstructure and sectioned according to different grains in the microstructure. Each section representing a different grain in the microstructure is assigned material properties based on their orientations. This creates the microstructure models before CMAS penetration (Supplementary Figure 2).

The microstructure models with CMAS penetration are generated by introducing CMAS layer along the grain boundaries. The grains are displaced to make space for CMAS by applying outward radial translation to their centroids with reference to the center of the entire microstructure. The outward radial translation is adjusted such that the overall empty spaces created along the grain boundaries reflect the desired level of CMAS penetration, *i.e.*, 3%, 5%, 8%, 11% and 15% by volume, into the microstructure. Similar to the microstructure models without CMAS penetration, a uniform mesh of 2D plane strain quadrilateral finite elements is laid and sectioned based on different microstructure regions that include grains and CMAS along the grain boundaries (Figures 4a and 4b).

**Cohesive Finite Element Model (CFEM)**

A cohesive finite element model (CFEM) with zero-thickness cohesive elements embedded along the edges of all the 2D plane strain quadrilateral elements (Figure 7a) [23], [24], [25] is employed to investigate the fracture characteristics of EBCs. The CFEM model is implemented in Abaqus® (Dassault Systemes Simulia Corporation, Providence, Rhode Island, USA) [40] and uses its standard bilinear traction separation law of the form,

$$\mathbf{t} = \begin{Bmatrix} t_n \\ t_s \end{Bmatrix} = (1-D) \begin{bmatrix} K_{nn} & 0 \\ 0 & K_{ss} \end{bmatrix} \begin{Bmatrix} \delta_n \\ \delta_s \end{Bmatrix}, \quad (1)$$

where $\mathbf{t}$ is the traction vector on the cohesive element, $D$ is scalar damage, $K$ is the cohesive stiffness of the element, and $\delta$ is the separation between its faces. A linear scalar damage law is defined as,

$$D = \frac{\delta_c(\delta - \delta_0)}{\delta(\delta_c - \delta_0)}, \quad (2)$$

where $\delta$ is the current effective separation, *i.e.*, $\delta = \sqrt{\delta_n^2 + \delta_s^2}$, $\delta_0$ is the initial separation for damage initiation, and $\delta_c$ is the critical separation at failure. The initial separation ($\delta_0$) is defined as the separation when the damage initiation criterion

$$\left\{ \left(\frac{t_n}{T_{max}}\right)^2 + \left(\frac{t_s}{T_{max}}\right)^2 \right\}^{0.5} = 1 \quad (3)$$

based on quadratic traction is satisfied. The maximum traction $T_{max}$ is assumed same for normal and shear directions [23]. The cohesive parameters used in this study are listed in Table 2.

A schematic diagram of the computational model used in CFEM simulations is shown in Figure 7b. A total of 32 CFEM simulations are performed by initiating crack along all four edges of each of the EBC microstructures obtained from eight EBSD images shown in Supplementary Figure 1. The cohesive element size is taken as 0.3μm to ensure convergence of fracture simulations for the given material properties and cohesive zone parameters [23], [41]. Surfing boundary conditions are used to induce stable crack propagation through the microstructures as shown in Figure 7b [33]. The *J*-integral is computed using a contour integral approach on a homogenous region surrounding the microstructures (Figure 7b) [33], [34], [42]. All the materials in this work are considered brittle and their material properties are given in Table 2. The computed *J*-integral is used to estimate the fracture toughness using the relation, $K_{Ic} = \sqrt{\frac{J\bar{E}}{1-\bar{v}}}$, where *J* is the computed *J*-integral, $\bar{E}$ is the effective Youngs modulus and $\bar{v}$ is the effective Poisson's ratio [30]. The homogeneous region surrounding the microstructure is assigned homogenized material properties of the microstructure it surrounds. The cohesive zone parameters based on bilinear traction separation laws are used for the cohesive elements inside the ceramic grains, *i.e.*, $LuPO_4$ and $Lu_2SiO_5$, and CMAS regions. For cohesive elements along the grain boundaries a sinusoidal fracture strength variation is assumed. It takes maximum and minimum value as 100% and 58% of the fracture strength of the grain, respectively [23]. The fracture strengths are assigned to the cohesive elements on grain boundaries based on misorientation between the adjacent grains [25]. The cohesive zone parameters are derived from the fracture toughness measured from indentation fracture experiments performed in this work. While the isotropic elastic properties are experimentally measured and the anisotropic elastic properties are taken from the literature [2], [36], [43]. The elastic properties of the homogenous regions surrounding CMAS-penetrated microstructures are computed using the rule of mixtures [33].

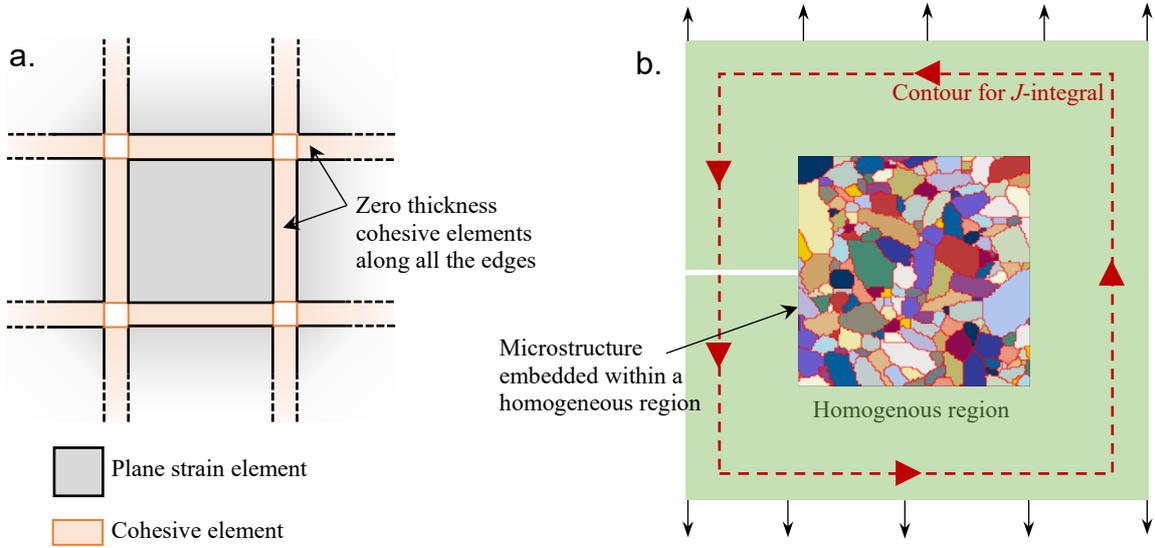

**Figure 7:** Schematic representation of the **(a)** cohesive finite element discretization and **(b)** computational model for CFEM simulations subjected to surfing boundary condition.

**Table 2:** Material properties and cohesive zone parameters.

| | Elastic Properties | | Cohesive Zone Parameters | | |
|---|---|---|---|---|---|
| | Isotropic | Anisotropic (GPa) | Cohesive Stiffness ($K_{nn} = K_{ss}$) | Critical Separation ($\delta_c$) | Max Traction ($T_{max}$) |
| LuPO$_4$ | $E$ = 193.4 GPa; $v$ = 0.29 | $C_{11}$ = 326.9; $C_{12}$ = 32.6; $C_{13}$ = 111; $C_{22}$ = 282.9; $C_{33}$ = 403.8; $C_{44}$ = 80.4; $C_{66}$ = 31.9 | 1000 N/μm² | 0.01 μm | 0.0016 N/μm² |
| Lu$_2$SiO$_5$ | $E$ = 148.6 GPa; $v$ = 0.24 | $C_{11}$ = 240; $C_{12}$ = 67; $C_{13}$ = 86; $C_{15}$ = 9; $C_{22}$ = 221; $C_{23}$ = 49; $C_{25}$ = -20; $C_{33}$ = 194; $C_{35}$ = -35; $C_{44}$ = 52; $C_{46}$ = 11; $C_{55}$ = 81; $C_{66}$ = 71 | 1000 N/μm² | 0.01 μm | 0.0015 N/μm² |
| CMAS | $E$ = 104.5 GPa; $v$ = 0.26 | --- | 1000 N/μm² | 0.01 μm | 0.0008 N/μm² |
| CMAS-grain Interface | --- | --- | 1000 N/μm² | 0.01 μm | 0.0004 N/μm² |
| Grain Boundaries | --- | --- | 1000 N/μm² | 0.01 μm | $T_{max}\times\{0.785+0.215\sin(4\theta)\}^*$ |

*$\theta$ - Average misorientation angle between adjacent grains

**Material Preparation and CMAS Reaction**

The EBCs pellets were sintered using spark plasma sintering (SPS) method. The LuPO$_4$ powder used in the experiments was synthesized as reported by Bryce *et al*. [32]. Lu$_2$SiO$_5$ was synthesized via a solid-state reaction, where the appropriate ratios of Lu$_2$O$_3$ and SiO$_2$ were milled together via high-energy ball milling (HEBM) at 500 rpm using ethanol as the mixing solvent and $ZrO_2$ milling balls as the grinding material. The model CMAS (33CaO-9MgO-13AlO1.5-45SiO2) is prepared by mixing oxide powders CaO (99.99%, Sigma-Aldrich), MgO (99.99%, Sigma-Aldrich), Al2O3 (99.99%, Sigma-Aldrich), and $SiO_2$ (99.9%, Alfa Aesar) according to the stoichiometric ratio. The mixture was then heated at the rate of 10 °C min$^{-1}$ up to 1300 °C for 8 hours and, finally, quenched in water to prepare the model CMAS. The $LuPO_4$-CMAS and $Lu_2SiO_5$-CMAS composites (with 5% and 8% CMAS) were also synthesized via solid-state ball milling of $LuPO_4$ and $Lu_2SiO_5$ powders with the model CMAS powder in their appropriate volume ratios. The milled powders were dried and then ground into fine powders, which were loaded into 10 mm diameter cylindrical graphite dies and sintered into dense pellets using SPS (Model 10−3 SPS system, Thermal Tech. LLC, Santa Rosa, CA). The powder samples were heated from room temperature to the sintering temperature of 1500 °C at a rate of 200 °C min$^{-1}$, then held at 1500 °C and under a uniaxial pressure of 50 MPa for 15 min. The sintered pellets were then ground with SiC abrasive papers and then polished with 0.06 μm silica colloid to achieve a mirror finish. The CMAS reactions on $LuPO_4$ and $Lu_2SiO_5$ were performed in a box furnace at 1300 °C with a heating rate of 10 °C min$^{-1}$ for 5 hours. The reacted samples were cut in half, processed, and polished for analysis.

**Measurement of Indentation Fracture Toughness**

Fracture toughness was estimated from a combination of micro indentation and nanoindentation measurements. A micro-hardness tester (Leco M-400) was used to create 10 indentations, each sample with a load of 1 kgf (~9.8 N), and each indentation lasted for 15 seconds. The crack lengths of these indentations were measured from the SEM image of each indentation. Nanoindentation tests were also carried out for each sample with a load of 1000 μN used to measure the hardness and elastic modulus of each sample. The fracture toughness ($K_c$) was evaluated using Eq (4),

$$K_c = \delta \left(\frac{E}{H}\right)^{0.5} \frac{P}{c^{1.5}}, \tag{4}$$

where $E$ is the elastic modulus, $H$ is the hardness, $P$ is the load for each micro indentation, $c$ is the average value of all crack lengths from the micro indentation test and $\delta$ is the indenter geometry related parameter.

**DATA AVAILABILITY**

The datasets/codes/scripts generated and/or analyzed during the current study are available from the corresponding author upon reasonable request.

## AUTHOR CONTRIBUTIONS

R.R., L.Z., L.H., J.L., and S.D. conceived the research. S.S. developed the methodology, implemented simulation models and analyzed the results. B.P.M. and K.B. synthesized and characterized the EBC samples and measured the material properties. R.R., L.Z., L.H., J.L., and S.D. supervised the project. All authors reviewed and edited the manuscript.

## ACKNOWLEDGEMENTS

This work was funded by the Division of Materials Research, National Science Foundation under Award DMREF-2119423. The funder played no role in study design, data collection, analysis and interpretation of data, or the writing of this manuscript.

## COMPETING INTERESTS

All authors declare no financial or non-financial competing interests.

# Supplementary Data

## 1. Electron back scattered diffraction (EBSD) micrographs

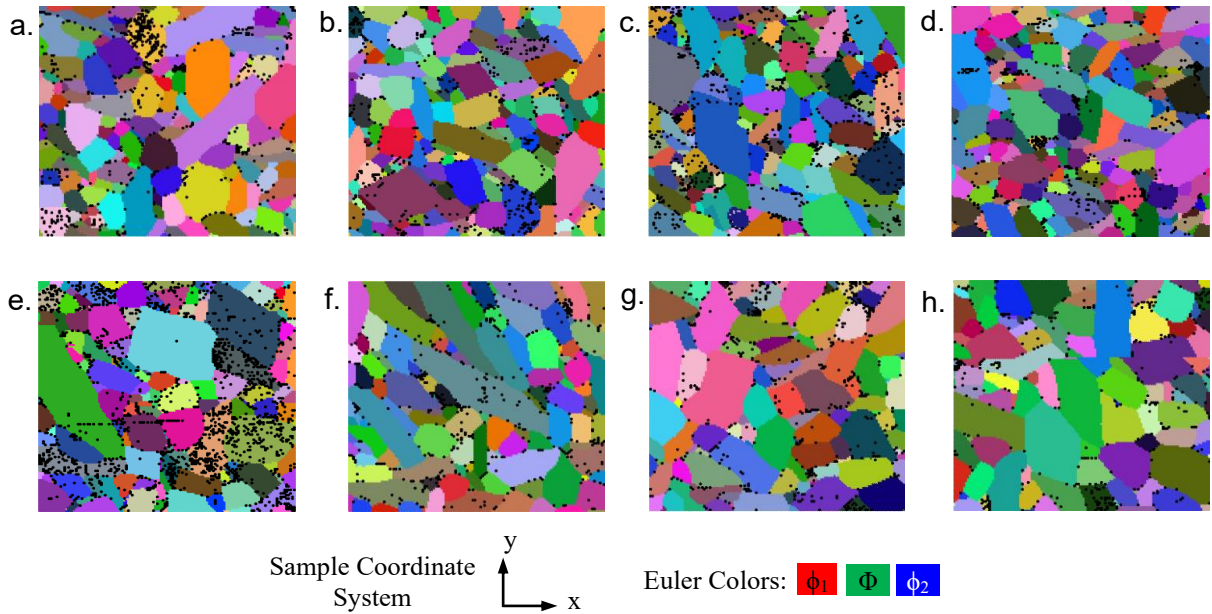

**Supplementary Figure 1: Electron backscattered diffraction (EBSD) micrographs obtained from the LuPO₄ sample.** Panels **(a)** through **(h)** show the raw EBSD micrographs. The orientations of each individual grain in micrographs are shown with reference to the sample coordinate system. Their respective colors represent the rendering based on the values of individual Euler angles.

## 2. Abaqus models from EBSD micrographs

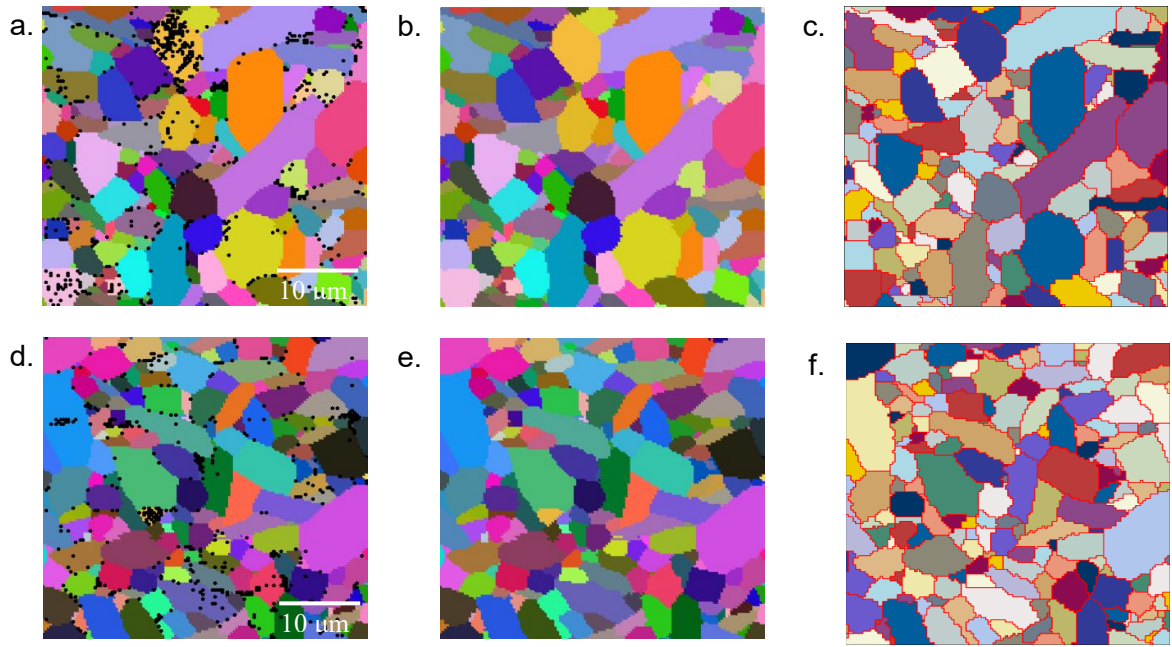

**Supplementary Figure 2: Generation of Abaqus model from electron backscatter diffraction (EBSD) micrographs.** Panels **(a)** and **(d)** show the raw EBSD micrographs. The raw EBSD micrographs are processed to remove non-indexed data in **(b)** and **(e)**. The grains and grain boundaries are assigned different sections in Abaqus, and material properties are assigned based on orientation and misorientation as shown in **(c)** and **(f)**.

## 3. Statistical Analysis

The flowchart of the statistical analysis to determine the similarity between simulation results and experimental measurements using univariate hypothesis testing is shown in Supplementary Figure 3. The null hypotheses for each of the hypothesis tests used in the statistical analysis are listed in Supplementary Table 1. Based on the univariate hypothesis tests, the calculated *p*-values are shown in Supplementary Tables 2 and 3 for pure and CMAS-penetrated materials, respectively. In Supplementary Tables 2 and 3, the boldface text represents the *p*-value of the hypothesis test that determines the similarity between simulations and experiments.

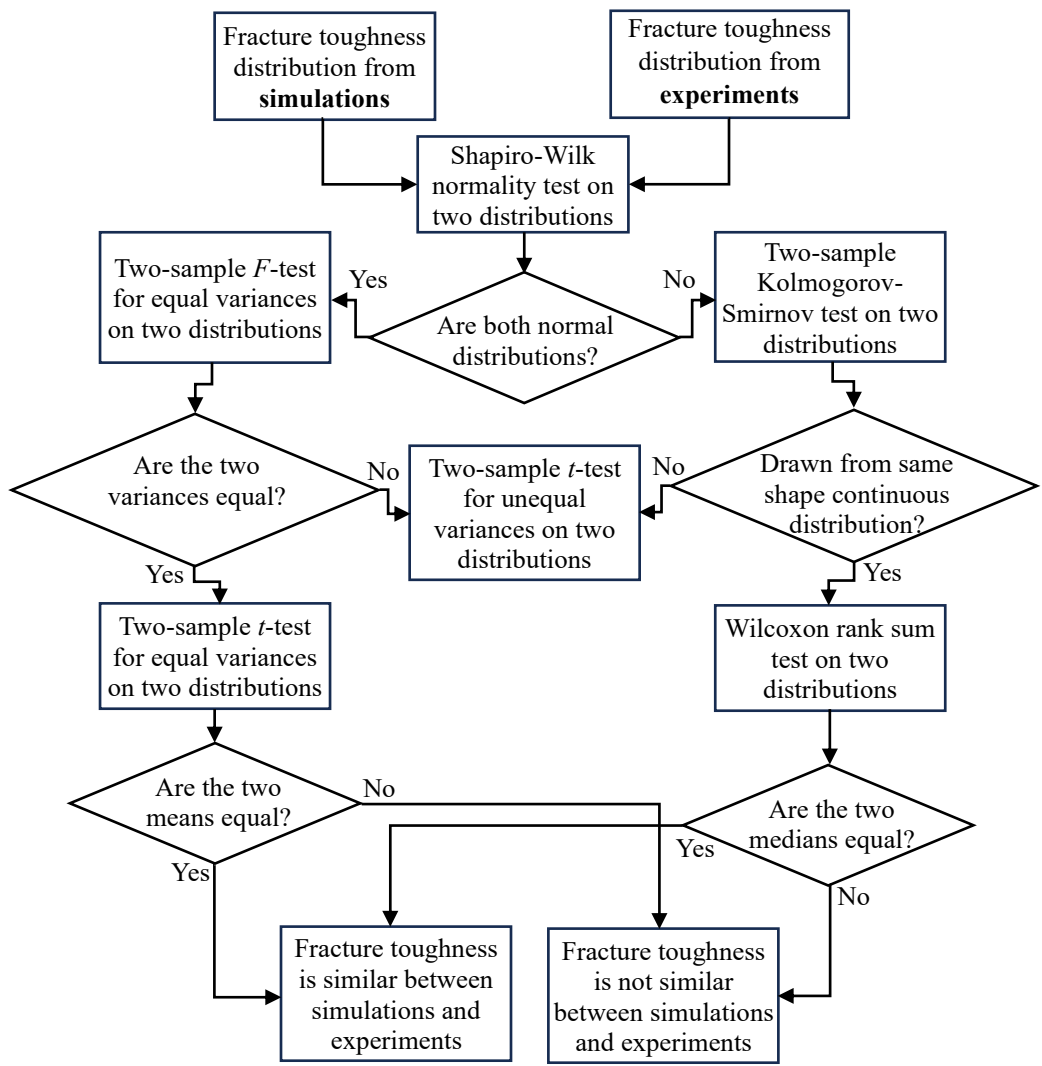

**Supplementary Figure 3: Statistical testing flow chart for comparing fracture toughness obtained from CFEM simulations and indentation experiments.**

**Supplementary Table 1**: Null hypotheses for the statistical tests used in this work

| Hypothesis test | Null hypothesis ($H_0$) |
|---|---|
| Shapiro-Wilk normality test | The data was drawn from a normal distribution |
| Two-sample $F$-test | Two data samples were drawn from normal distributions of same variance |
| Two-sample Kolmogorov-Smirnov test | Two data samples were independently drawn from distributions that have the same shape but may have a different mean |
| Two-sample $t$-test (Equal variances) | Two data samples were drawn independently from two normal distributions with equal means but unknown and equal variances |
| Two-sample $t$-test (Unequal variances) | Two data samples were drawn independently from normal distributions with equal means but unknown and unequal variances |
| Wilcoxon rank-sum test | Two data samples were drawn from continuous distributions with the same shape and equal medians |

**Supplementary Table 2**: The $p$-values of various statistical tests used to assess similarity ($p < 0.01$) of experiment and simulation data for pure materials.

| Material → | | Pure LuPO$_4$ | | Pure Lu$_2$SiO$_5$ | | Pure CMAS | |
|---|---|---|---|---|---|---|---|
| Hypothesis tests ↓ | | Exp | CFEM | Exp | CFEM | Exp | CFEM |
| Shapiro-Wilk normality test | $p$-value | 0.2182 | 0.0099 | 0.8412 | 0.3523 | 0.4956 | 0.1847 |
| | Normal distribution? | Yes | No | Yes | Yes | Yes | Yes |
| Two-sample $F$-test | $p$-value | --- | | 0.9015 | | 0.000 | |
| | Equal variance? | --- | | Yes | | No | |
| Kolmogorov-Smirnov test | $p$-value | 0.1306 | | --- | | --- | |
| | Same distribution? | Yes | | --- | | --- | |
| Two-sample $t$-test (Equal variance) | $p$-value | --- | | **0.1063** | | --- | |
| | Equal means? | --- | | **Yes** | | --- | |
| Two-sample $t$-test (Unequal variance) | $p$-value | --- | | --- | | **0.8857** | |
| | Equal means? | --- | | --- | | **Yes** | |
| Wilcoxon rank-sum test | $p$-value | **0.2095** | | --- | | --- | |
| | Equal medians? | **Yes** | | --- | | --- | |
| Outcome | | **Similar** | | **Similar** | | **Similar** | |

**Supplementary Table 3:** The *p*-values of various statistical tests used to assess similarity ($p < 0.01$) of experiment and simulation data for CMAS-penetrated materials.

| Material → | | LuPO$_4$ + 5% CMAS | | LuPO$_4$ + 8% CMAS | | Lu$_2$SiO$_5$ + 5% CMAS | | Lu$_2$SiO$_5$ + 8% CMAS | |
|---|---|---|---|---|---|---|---|---|---|
| Hypothesis tests ↓ | | Exp | CFEM | Exp | CFEM | Exp | CFEM | Exp | CFEM |
| Shapiro-Wilk normality test | *p*-value | 0.2250 | 0.0042 | 0.2435 | 0.0412 | 0.1367 | 0.0022 | 0.4521 | 0.0574 |
| | Normal distribution? | Yes | No | Yes | Yes | Yes | No | Yes | Yes |
| Two-sample *F*-test | *p*-value | --- | --- | 0.5329 | | --- | --- | 0.4204 | |
| | Equal variance? | --- | --- | Yes | | --- | --- | Yes | |
| Kolmogorov-Smirnov test | *p*-value | 0.0111 | | --- | --- | 0.3516 | | --- | --- |
| | Same distribution? | Yes | | --- | --- | Yes | | --- | --- |
| Two-sample *t*-test (Equal variance) | *p*-value | --- | --- | **0.006** | | --- | --- | **0.1069** | |
| | Equal means? | --- | --- | **No** | | --- | --- | **Yes** | |
| Two-sample *t*-test (Unequal variance) | *p*-value | --- | --- | --- | --- | --- | --- | --- | --- |
| | Equal means? | --- | --- | --- | --- | --- | --- | --- | --- |
| Wilcoxon rank-sum test | *p*-value | **0.2073** | | --- | --- | **0.8018** | | --- | --- |
| | Equal medians? | **Yes** | | --- | --- | **Yes** | | --- | --- |
| Outcome | | **Similar** | | **Not Similar** | | **Similar** | | **Similar** | |